\documentclass[aps,prl,twocolumn,notitlepage,superscriptaddress,floatfix]{revtex4-1}

\usepackage[T1]{fontenc} 

\usepackage[unicode,colorlinks]{hyperref}
\usepackage{amsmath,amssymb,amsfonts}
\usepackage{bbm}         
\usepackage{graphicx}
\usepackage{physics}

\usepackage[dvipsnames]{xcolor}

\renewcommand{\subsubsection}[1]{\teal{#1.}}
\newcommand{\teal}[1]{\textcolor{teal}{\textit{#1}}}

\renewcommand{\subsection}[1]{\teal{#1.---}}
\renewcommand{\section}[1]{\teal{#1.---}}

\begin{document}

\title{Flux superperiods and periodicity transitions in quantum Hall interferometers
}

\author{Bernd Rosenow}
\affiliation{Institut f\"{u}r Theoretische Physik, Universit\"{a}t Leipzig, Br\"{u}derstrasse 16, 04103 Leipzig, Germany}
\affiliation{Department of Condensed Matter Physics, Weizmann Institute of Science, Rehovot 76100, Israel}
\author{Ady Stern}
\affiliation{Department of Condensed Matter Physics, Weizmann Institute of Science, Rehovot 76100, Israel}
\date{July 14 2019}

\begin{abstract}
\normalsize

For strongly screened Coulomb interactions, quantum Hall interferometers can operate in a novel regime: the intrinsic energy gap
can be larger than the charging energy, and addition of flux quanta can occur without adding quasi-particles. We show that flux superperiods are possible, and  reconcile their appearance with  the Byers-Yang theorem. We  explain that the observation of anyonic statistical phases is possible by tuning to the transition from a regime with constant chemical potential to a regime
with constant particle density, where a
flux superperiod changes to a periodicity with one flux quantum at a critical magnetic field strength.
\end{abstract}

\maketitle

%

Fractionally charged quasi-particles (qps) are among the most intriguing aspects of the fractional quantum Hall effect. Their existence and their fractional anyonic statistics are believed to be unavoidable consequences of the quantization of the Hall conductivity to a fractional value of $e^2/h$, with $e$ the electron charge, and $h$ Planck's constant.

Much experimental effort has been devoted towards a measurement of the fractional charge and the unconventional statistics. The fractional charge has been measured through noise \cite{shot1,shot2,shot3,Dolev+08} and electrometry \cite{Goldmann95,Martin+04,Vivek+11}. For the statistics, defined by the geometric phase obtained when one qp encircles another, the natural tool to use is interferometry \cite{Chamon+97}. 
Indeed, many attempts to study integer and fractional quantum Hall states through Fabry-Perot 
\cite{Zhang+09,Lin+09,Ofek+10,McClure+09,McClure+12,Choi+11,Baer+13,Camino+05,Camino+07,Willett+09,Willett+13}
and Mach-Zehnder interferometry \cite{Ji+03,Roulleau+08,Gurman+16}      have been carried out.

Interpreting interference experiments in quantum Hall (QH) Fabry-Perot devices was found to be complicated by the strong Coulomb interaction, which
can drive the interferometer into a Coulomb-dominated regime \cite{subperiod,FPtheory}. This problem is particularly severe in small (micron size) interferometers, which are favored from the point of view of battling decoherence. Attempts to screen the Coulomb interaction using Ohmic contacts situated inside or close to a Fabry-Perot interferometer were successful in the integer regime, resulting in
novel halving of flux and gate voltage periodicities \cite{Choi+15,Frigeri+17}.

More recently, the long-range Coulomb coupling has been addressed by the introduction of screening layers both above and below the transport layer \cite{manfra19}. In this setup, the phenomenology of the weakly interacting Aharonov-Bohm (AB) regime was observed, and an interference signal for the fractional $1/3$ state was reported. Rather unexpectedly, the oscillating part of the conductance through the interferometer followed a flux "superperiod" of
three flux quanta $\Delta\Phi=3\Phi_0= 3 h/e$, as compared to the previously predicted period, $\Phi_0$ \cite{FPtheory} (for a discussion of  superperiods in other  geometries see \cite{ThouGe,GeThou,Jain+93}). Since the interior of the interferometer is incompressible and not accessible to the interfering qps, one may wonder whether such a super-period violates the Byers-Yang theorem \cite{ByYa61}, which states that in the presence of a "hole" in a multiply connected system, all observables are periodic when the magnetic flux through the hole is adiabatically varied by $\Phi_0$.

%
\begin{figure}
\includegraphics[width=0.45\textwidth]{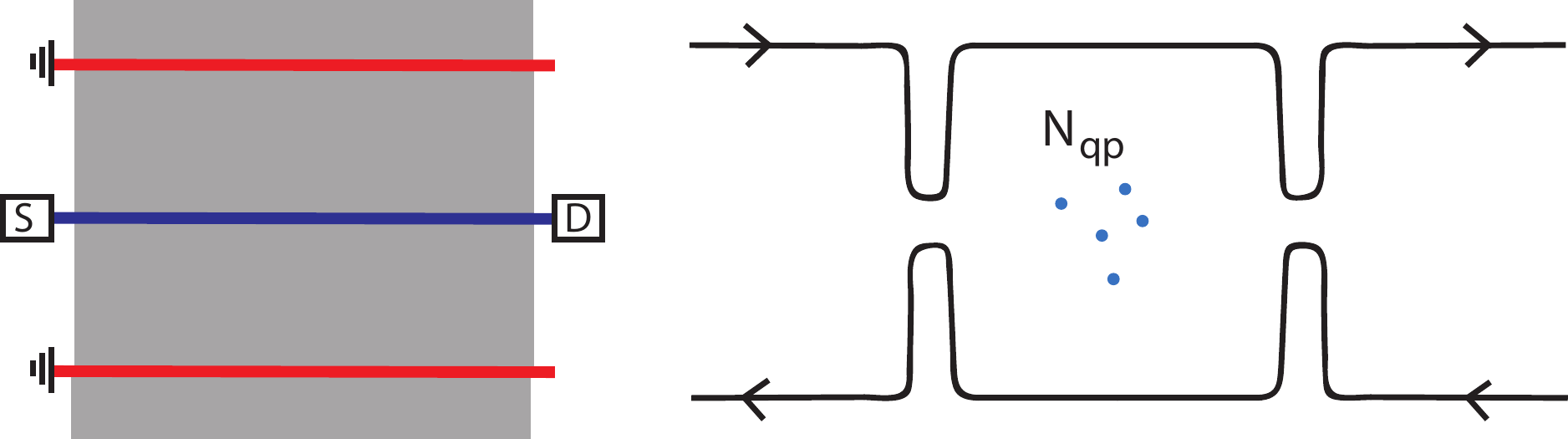}%
\caption{%
Left panel: Side view - schematic heterostructure with a transport layer (blue) and two screening layers, separated by insulating barriers (grey). Right panel: Top view -  Fabry-Perot interferometer with $N_{qp}$ qps localized in the interference cell.
\label{fig-setup}}
\end{figure}

In this letter, we formulate a theory of QH Fabry-Perot interferometers in the presence of screened Coulomb interaction. Taking into account the competition between the charging energy and the energy gap for  qp excitations,
we find that around the center of QH plateaus  there is a region of magnetic field $B$ in which small variations of $B$ does not lead to  qp addition in the bulk. The width of this region sets a lower bound for the extension of QH plateaus. For an interferometer, this implies that there
is a region of $B$ in which the number of qps in the interference loop is $B$--independent. When that happens, the interfering qp accumulates
an AB phase which is reduced compared to the electronic one by the ratio of qp charge $e^*$ and electron charge $e$, hence leading to a magnetic flux superperiod. Beyond the critical magnetic field limiting this region, every additional flux quantum causes the creation of bulk qp(s), augmenting the AB phase by an anyonic phase and reducing the the interference period to $\Phi_0$.

Before discussing interferometry, we consider a bulk transport layer with particle density $n_1$ capacitively coupled to a screening layer with density $n_2$. We assume no tunneling between the layers. Coulomb coupling between the two layers creates image charges in the screening layer, making the
 interaction in the transport layer short-ranged. Viewing the combination of the two layers as a planar capacitor, the
charging of the capacitor can be described by a density variation $\delta n$, such that  the density of the transport layer is $n_1 = {\nu\over 2 \pi \ell_{B_0}^2}  + \delta n$ and the density
of the screening layer is $n_2 = n_2^{(0)} - \delta n$. Here, $B_0$ is the reference magnetic field in the center of the quantum Hall plateau,
and $\ell_B = \sqrt{\hbar/e B}$ is the magnetic length.

We consider an incompressible QH state, with a possible finite density of
qps if the magnetic field is lowered as compared to $B_0$. Then, the energy density is  given by
%
\begin{eqnarray}
E & = & \left(E_0 - \mu \right)  {\nu \over 2 \pi \ell_B^2} + \Delta_{e^*}(B){1 \over e^*}  \left({\delta n} + { \nu\over 2 \pi \ell_{B_0}^2}
- {\nu \over 2 \pi \ell_B^2}\right)\nonumber \\
& & + {\delta n^2 e^2\over 2} \left({1 \over C_g} + {1 \over C_q}\right) \ \ .
\label{energydensity.eq}
\end{eqnarray}
%

Here, $E_0$ is the ground state energy per electron in the pristine QH state with filling factor $\nu$, and $e^*$ is measured in units of the electron charge.
Furthermore, $\Delta_{e^*}(B)$ is the magnetic field dependent energy gap for adding a qp,  $C_g$ denotes the geometric capacitance between transport and screening layer, and $C_q$ is the quantum capacitance of the screening layer.
In the second term on the r.h.s.
of Eq.~(\ref{energydensity.eq}), the overall factor of $1/e^*$ is due to the ratio of the density of qps to the density of electrons, and the
factors $\nu$ combined with the area per electron account for the filling of the Landau level.  The extra filling above the quantized fraction is
%
\begin{equation}
\delta \nu \ = \ 2 \pi \ell_B^2 \delta n \ + \  \nu {\ell_B^2 \over \ell_{B_{0}}^2 } \ - \ \nu \ \ .
\end{equation}
%
When minimizing the energy density Eq.(\ref{energydensity.eq}) with respect to $\delta n$, we find
%
\begin{equation}
\delta n  \ = \ - {{\Delta_{e^*}(B)} \over e^* e^2 \left({1 \over C_{\rm g}} + {1 \over C_q} \right) } \ \ .
\label{density_fractional.eq}
\end{equation}
%

According to our assumption $B < B_0$, the density of qps  cannot be negative, and we find the constraint
%
\begin{equation}
\delta n   + { \nu \over  2 \pi \ell_{B_0}^2}
- {\nu \over  2 \pi \ell_B^2}  \  \geq \ 0  \ \ .
\label{inequality_fractional.eq}
\end{equation}
%
Clearly, for small changes of $B$ with $(B_0 - B)/B_0 \ll1$ the density change found in Eq.~(\ref{density_fractional.eq}) violates
the above inequality. Therefore, the actual density change is determined by viewing Eq.~(\ref{inequality_fractional.eq}) as an equality, which implies that $\delta\nu=0$, i.e., the filling factor stays constant while $B$ is varied away from $B_0$.
Thus,
%
\begin{equation}
\delta n \ = \ - \nu {B_0 - B \over \Phi_0}  \ ,  \ \ \  B_c < B < B_0 \ \ ,
\label{deltan_inequality.eq}
\end{equation}
%
When $B$ is further lowered away from $B_0$, a critical magnetic field $B_c$ is reached at which
the expressions for the density change Eqs.~(\ref{density_fractional.eq}), (\ref{deltan_inequality.eq}) are equal. From this condition, we
obtain
%
\begin{equation}
 { B_c } \ = \   {B_0 } \ - \ {\Delta_{e^*}(B) \Phi_0 \over \nu e^*  \left({e^2 \over C_{g}} + {e^2 \over C_q}\right)}  \ \ .
 \label{criticalfield.eq}
\end{equation}
%
For $B < B_c$, the filling factor does not remain constant anymore, the density change is determined by Eq.~(\ref{density_fractional.eq}), and its $B$--dependence is determined by the $B$--dependence of the energy gap $\Delta_{e^*}$.

For integer QH states $e^*=1$, and the energy gap is given by either the exchange enhanced spin splitting, or the cyclotron gap, depending on whether the lowest unoccupied Landau level has the same orbital index as the highest occupied one. For concreteness, we discuss a gap
given by the cyclotron energy $\hbar e B/m^*$, where $m^*$ denotes the effective mass of electrons. Then,
%
\begin{equation}
B_{c,\rm int} \left( 1 + {2 \pi  \hbar^2  C_{\rm tot} \over e^2 m_{\rm eff}}\right) \ = \ B_0 \ \ ,
\end{equation}
%
where  ${1 \over C_{\rm tot}} \equiv  {1 \over C_g} + {1 \over C_q}$. We note that usually the geometric capacitance dominates over the quantum capacitance, such that 
%
\begin{equation}
{1 \over C_{\rm tot}} \ \gtrapprox   \ {d  \over \epsilon }  \ \ .
\end{equation}
%
With one screening layer, $d$ denotes the spatial distance between transport layer and screening layer. With  screening layers on both sides of the transport layer, the capacitances  add up and $d$ is half of this distance. In the experiment reported in \cite{manfra19}, there are two such screening layers, each about $50 nm$ from the transport layer, and $d \approx 25 nm$  \cite{manfra19}. Introducing the Bohr radius $a_0 = {4 \pi \epsilon_0 \hbar^2 \over m_e e^2} $, we express the critical magnetic field as
%
\begin{equation}
B_{c, \rm int} \ = \ B_0 \left(1 \ + \ {\epsilon_r \over 2}  {m_e \over m_{\rm eff}} \, {a_0 \over d} \right)^{-1} \ \ .
\label{bc_integer.eq}
\end{equation}
%
For the experimentally relevant parameters $m_{\rm eff} = 0.07 m_e$, $\epsilon_r = 13$, $d = 25 nm$ we find that $B_{c, \rm int}
\approx {5 \over 6} B_0$, implying that the total magnetic field width of the $\nu=2$ plateau should be at least $1/3$ of the magnetic field at its center, in agreement with the plateau width in \cite{manfra19}.

 Turning now to a Fabry-Perot interferometer with an interference cell of radius $R$, we note that the screening layer(s) significantly weakens the interaction coupling between the edge and bulk of an interferometer: only an annulus of bulk area
$2\pi R d \ll \pi R^2$ interacts with the edge. Since only a small part of charges accumulated in the bulk interacts with the edge, one can expect that the bulk-edge coupling is parametrically smaller than the self-interaction of edge charges, placing the interferometer in the AB regime \cite{subperiod,FPtheory}. It is indeed this regime which was observed in 
Ref.~\cite{manfra19}.  Since there are no anyonic phases for integer QH states the flux periodicity is not affected by whether or not electrons enter the interference cell.  For filling fractions higher than two however, a sufficiently strong  inter-mode Coulomb coupling, together with a weak bulk-edge coupling, halves both flux and gate voltage periods
\cite{Frigeri+17}. Observation of such a halving in the experiment \cite{manfra19} indicates that the above conditions are indeed satisfied.

For the fractional QH effect the energy gap is set by the Coulomb interaction    $\Delta_{e^*} = \alpha_{e^*} {e^2 \over 4 \pi \epsilon \ell_B}$,
with $0.01 \lesssim \alpha \lesssim 0.05$  parametrizing the specific value of the energy gap for different fractional quantum Hall states \cite{Morf03}.
With this, we obtain
%
\begin{eqnarray}
B_{c,\rm frac}  & = &   B_0 \left( 1 \ + \ {\gamma^2 \over 2} \ - \ \gamma \sqrt{1 + {\gamma^2 \over 4} }\right)
\label{bc_fractional.eq}
 \end{eqnarray}
 %
with
%
\begin{equation}
\gamma \  = \ {\alpha_{e^*} \over 2 \nu e^*} \, {\ell_{B_0} \over d}  \ \ .
\end{equation}
%
For a gap of $7K$ \cite{manfra19} for $\nu=1/3$
we obtain $\gamma \approx 0.018$.  As a consequence, the relative change
in $B$ before qps enter in the bulk is about 2\%.

\begin{figure}
\includegraphics[width=0.4\textwidth]{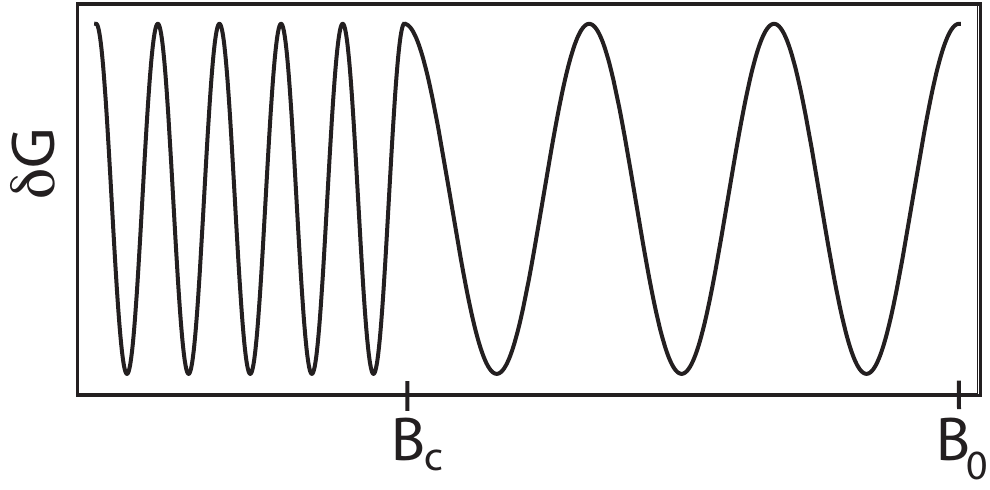}%
\caption{%
Change in oscillation periodicity of a Fabry-Perot interferometer as a function of magentic field for the $\nu=1/3$ quantum Hall state. For magnetic fields near the plateau center with $B_c < B < B_0$, no qps enter the interference cell and the magnetic fiel period is
$\Delta B = 3 \Phi_0/A$. For $B < B_c $, a quasi-hole enters for each additional flux quantum, and the period changes to $\Delta B =  \Phi_0/A$ due to the combination of Aharonov-Bohm phase and anyonic statistical phase.
\label{fig-oscillations}}
\end{figure}

Using the expressions we derived for the bulk density variation with $B$, we may estimate the number of qps inside an interferometer of area $A$. As long as that number does not change with the magnetic field, the interference follows a superperiod. We find the interferometer to host
%
\begin{eqnarray}
 N_{qp} & = &
 {A(B_0 - B) \over  \Phi_0}\  -  \   { 1 \over  2 \nu e^*} \,   {\Delta_{e^*} \over E_c}  \ , \ \ \  B < B_c
  \end{eqnarray}
%
qps, where the charging energy is $E_c = {e^2  d \over 2 \epsilon A}$. Using the same parameters as above, the area per flux quantum is $2 \pi \ell_B^2 \approx 6 \cdot 49 nm^2 \approx 300 nm^2$. The area of the interferometer \cite{manfra19} can be estimated from the magnetic field periodicity $\delta B \approx 7 mT$ in the integer regime according to $A = \phi_0 / \delta B \approx 0.56 \mu m^2$.
For this area, there
would be approximately $1.87 \cdot 10^3$ flux quanta in the interferometer cell, and the offset in the qp number  would be $33$ for a gap of $2K$ and $117$ for a gap of $7K$, i.e.~there would  be
$11-39$ oscillations with periodicity $3 \Phi_0$ before extra qps enter. After that, the periodicity should be reduced to $\Phi_0$, see the discussion below.
We get a similar estimate by using the experimental values \cite{manfra19} for $\Delta_{e^*} \approx 700 \mu eV$ and the charging energy
$E_c \approx 17 \mu eV$ to obtain an offset of $700/(2\cdot 17/9)\approx 126$ flux quanta through the interferometer, corresponding to 42 oscillations.



Focusing on $\nu=1/3$, we denote the range $B_c < B < B_0$ as regime I, and the range of $B < B_c$ but with the system still staying on the $\nu=1/3$ plateau as regime II. The range of the Coulomb interaction is determined by the distance $d$ between the transport and screening layer. As long as $d\gg l_B$, the short range part of the Coulomb interaction, which is responsible for the formation of fractional quantum Hall states, is unchanged, and one expects to find a stable hierarchy of fractional quantum Hall states.
For $\nu=1/3$ Laughlin states in regime I the $B$--dependence of the interference phase is purely due to the Aharonov-Bohm phase $e^* A B/\Phi_0$. As a consequence, the flux periodicity is
 $\Phi_0/e^*$, giving rise to a flux superperiod. Similarly, for $\nu = 1/(2 m +1)$ superperiods with $e^*=1/(2m+1)$ occur in regime I.

In contrast, in regime II a reduction of the magnetic flux by $\Phi_0$ introduces a qp in the interference loop. Then, the phase accumulated due to a flux change of one $\Phi_0$ is the sum of the AB phase $2 \pi e^*$ and a statistical phase $4 \pi  e^*$, which in sum yield
 a phase change of $2 \pi$, such that the flux period is $\Phi_0$.  A similar outcome holds for $\nu=1/(2m+1)$. Experimental observation of the phase shift $4 \pi  m e^*$
as a jump in the interference pattern would be a highly desirable demonstration of anyonic statistics. At finite temperature however, the
phase jump is smeared due to fluctuations in the qp number \cite{FPtheory} if the characteristic relaxation time is shorter than the measurement time. When Fourier transforming a phase jump, higher harmonics in the magnetic flux dependence arise. The  amplitude of the
$n$-th harmonic is proportional to $\exp\left[-\pi^2 (n-1+2 e^* m)^2  k_B T/E_c (e^*)^2\right]$ \cite{FPtheory}. For this reason, the temperature needs to satisfy $k_B T \leq (e^*)^2 E_c/\pi^2$ for phase jumps to be observable. While for the sample used in \cite{manfra19} this would necessitate
a temperature of $2 mK$, a moderate increase of the charging energy could make the observation of phase jumps feasible.

 As an additional experimental parameter, we consider  an external gate which changes the area of the interference cell according to
 $\delta A = {d A \over d V_G}  \delta V_G$, where ${d A \over d V_G}$ is assumed to depend only weakly on magnetic field.
 Such an area gate does not add qps to the bulk of the interferometer, so the gate voltage periodicity will be the same in regimes I and II. Then,   the gate voltage periodicity is given by $\Delta V_G =  \Phi_0/(e^* B {d A \over d V_G})$. Due to the fact that the magnetic field $B_{1/3}$ at the filling fraction $\nu =1/3$ is three times larger than the field $B_1$ at $\nu=1$, we find that the gate voltage periodicities are expected to be equal in the two cases, as observed in Ref. \cite{manfra19}.

  The Byers-Yang theorem \cite{ByYa61} states that in a multiply connected geometry with flux through a "hole",  all physical properties are periodic under a change of the magnetic  flux by one flux quantum. In the context of QH Fabry-Perot interferometers, there is no actual  "hole" through the interferometer, unless an anti-dot is placed inside the interference cell. However, the gapped bulk is inaccessible to interfering qps, and it is interesting to ask under which conditions the Byers-Yang theorem applies. In QH  interferometers, the interference phase is determined by the magnetic flux enclosed by the interference path. In a pristine fractional QH state with filling fraction $\nu$, the flux is tied to the number of electrons $N_{el} = \nu B A/\Phi_0$ encircled by the interfering electron \cite{Arovas}, and the interference phase can be interpreted as a statistical phase.
 In the limit of weak backscattering, the area $A$ encircled by the interfering particle is fixed due to electrostatic constraints, and
 in regime I with fixed chemical potential,  the filling fraction $\nu$ is independent of magnetic field. For these reasons,
  the number of encircled electrons is directly proportional to  magnetic field times filling fraction, and a super-period arises.
Such a change in the number of encircled electrons however would not be possible if there was a hole without electrons (an anti-dot) inside the interferometer.
In this situation,
only
 electrons in the narrow  annulus defined by the interfering edge on the outside and the perimeter of the anti-dot on the inside would contribute a statistical phase. If the area of this annulus is much smaller than the total interferometer area, then the
number of electrons inside the annulus is approximately independent of magnetic field. The period of one flux quantum would then arise due to the modulation of the anti-dot energy spectrum with magnetic flux, which indeed is governed by the prediction of the Byers-Yang theorem.

For filling factor $\nu=5/2$ interferometers are expected to be able to distinguish between abelian and non-abelian candidate states  \cite{Stern+2010} through the even-odd effect \cite{evenodd1,evenodd2,evenodd3}.
The magnetic field period for interfering non-abelian charges $e^* = {1 \over 4}$ qps is
 $4 \Phi_0$ if there is an even number of qps in the bulk of the interference cell, and $2 \Phi_0$ for an odd number of qps inside the interference cell. In addition, for an even number of qps, the interference phase can be shifted by $\pi$ depending on the internal state of the non-abelian degree of freedom of these qps. In regime I, the number of qps in the interior is independent of mangetic field, and is determined by the state of the pristine interferometer in the center of the plateau. Depending on the parity of the number of trapped qps in the bulk, the flux period could be $4 \Phi_0$ or $2 \Phi_0$.
 In regime II, a change of magnetic field is accompanied by a change of the number of qps in the bulk. Depending on the dynamics of the non-abelian degree of freedom associated with the bulk qps (the so-called neutral fermion), the
 flux period in the Pfaffian state (anti-Pfaffian state) for weak bulk-edge coupling can be $1 \Phi_0$ ($1.5 \Phi_0$) for fixed fermion parity, and $2 \Phi_0$ ($3 \Phi_0$)
 for random fermion parity, see \cite{keyserlingk+15}.

For $\nu=2/3$, the magnetic field periodicity in regime II in the limit of a closed interferometer was discussed in
\cite{Viola+12,PaGeSi15}. Due to the presence of a neutral mode \cite{KaFiPo94}, pairs of conductance peaks bunch together, giving rise to a doubling of the magnetic field period of the conductance as compared to a situation with a charge mode only.   Here we discuss an interferometer with weak backscattering, operating  in regime I. Generally, the charge and neutral modes have different velocities, $v_n,v_c$.   When $v_n\ll v_c$, the charge $e^* = 2/3$ quasiparticle will have the highest visibility since its tunneling operator does not excite the neutral mode. In regime I, this implies a flux period of $\phi_0/e^* = {3 / 2}$, somewhat larger than the experimentally observed value \cite{manfra19} of $ \phi_0$.
In contrast, if $v_n$ is sufficiently high
the interference of charge $e^*=1/3$ qps will be observable and contribute to the interference signal, with a flux period of $3 \phi_0$. In regime II, the statistical phase due to addition of qps in the interferometer cell will again reduce the flux period. In the limit of a slow neutral mode, a subperiod of $\phi_0/2$ is expected \cite{PaGeSi15}, which is doubled to $\phi_0$ for the case of $v_n\approx v_c$ \cite{Viola+12}.

Our analysis of the critical field $B_c$  assumed that the pristine QH state is strictly incompressible and has a finite gap for excitations. In reality however, disorder  will give rise to a finite density of states in the transport layer, which we parametrize with the help of a quantum capacitance $C_{q,t}$. Then, the second term Eq.~(\ref{energydensity.eq}) is replaced by
${e^2 \over 2 C_{q,t}}  \left( \delta n + {\nu \over 2 \pi \ell_{B_0}^2} - { \nu \over 2 \pi \ell_B^2} \right)^2$. Minimizing this
modified energy, one finds that the change in density is
%
\begin{equation}
\delta n \ = - \ { {\nu \over 2 \pi \ell_{B_0}^2} - { \nu \over 2 \pi \ell_B^2} \over 1 + C_{q,t} \left({1 \over C_g} + {1 \over C_q}\right) }  \ \ ,
\end{equation}
%
slightly smaller that the density change implied from  (\ref{inequality_fractional.eq}). Assuming the small density of states in the transport layer is independent of energy, the behavior described above continues until
the energy of excited states is lowered to the level of the chemical potential. This reduces $B_c$ to
%
\begin{equation}
B_{c,\rm dis} \ = \ B_c \left[ 1 + C_{q,t}\left({1 \over C_g} + {1 \over C_q}\right)\right]^{-1} \ \ ,
\end{equation}
%
When the states below the gap are localized and do not contribute to transport, disorder affects the bulk plateaus only by slightly increasing their width as compared to the pristine case.

Inside the interferometer, excited states occur at discrete energies smaller than the energy gap, with a spectrum that varies with disorder realization. Due to these states, qps will enter the interference cell at discrete values of the magnetic field, below the critical field.
For fractional interferometers, this will lead to phase shifts in the interference pattern, which allow to determine the anyonic statistical phase.

In summary, we presented here conditions under which a quantum Hall Fabry-Perot interferometer may exhibit a flux periodicity larger than a single $\phi_0$. We related these conditions to situations in which the quenching of charging energy makes the interferometer follow a line of constant filling factor, rather than constant density, when the magnetic field is varied.

\begin{acknowledgments}
We would like to thank M.~Manfra for a helpful discussion. B.R. acknowledges support  from the Rosi and Max Varon Visiting Professorship at the Weizmann Institute of Science, and from DFG grant RO 2247/8-1.
A.S. was supported by  CRC 183 of DFG, the Israel Science Foundation, and the European Research Council (Project LEGOTOP).
\end{acknowledgments}

\end{document}